\documentclass[fleqn, 10pt]{wlscirep}

\usepackage{mathpazo}
\usepackage{blkarray}

\definecolor{darkgreen}{RGB}{0,153,0}

\title{Lagrangian dynamical geography of the Gulf of Mexico}

\author[1,*]{P.\ Miron}
\author[1]{F.\ J.\ Beron-Vera}
\author[2]{M.\ J.\ Olascoaga}
\author[3]{J.\ Sheinbaum}
\author[3]{P.\ P\'erez-Brunius}
\author[4]{G.\ Froyland}

\affil[1]{Department of Atmospheric Sciences, Rosenstiel School of
Marine and Atmospheric Science, University of Miami, Miami, Florida,
USA} \affil[2]{Department of Ocean Sciences, Rosenstiel School of
Marine and Atmospheric Science, University of Miami, Miami, Florida,
USA} \affil[3]{Departamento de Oceanograf\'{\i}a F\'{\i}sica, Centro
de Investigaci\'on Cient\'{\i}fica y Educaci\'on Superior de Ensenada,
Ensenada, Baja California, Mexico} \affil[4]{School of Mathematics
and Statistics, University of New South Wales, Sydney, Australia}
\affil[*]{pmiron@rsmas.miami.edu}


\begin{abstract}
  We construct a Markov-chain representation of the surface-ocean
  Lagrangian dynamics in a region occupied by the Gulf of Mexico
  (GoM) and adjacent portions of the Caribbean Sea and North Atlantic
  using satellite-tracked drifter trajectory data, the largest
  collection so far considered.  From the analysis of the eigenvectors
  of the transition matrix associated with the chain, we identify
  almost-invariant attracting sets and their basins of attraction.
  With this information we decompose the GoM's geography into weakly
  dynamically interacting provinces, which constrain the connectivity
  between distant locations within the GoM.   Offshore oil exploration,
  oil spill contingency planning, and fish larval connectivity
  assessment are among the many activities that can benefit from
  the dynamical information carried in the geography constructed
  here.
\end{abstract}

\begin{document}

\flushbottom
\maketitle
\thispagestyle{empty}

\section*{Introduction}

Over the past few decades a number of satellite-tracked surface
drifting buoys have surveyed the Gulf of Mexico (GoM).  Much insight
into the GoM's surface-ocean Lagrangian dynamics has been gained
from the analysis of different subsets of the collected drifter
data.  A large body of the work done was dedicated to study relative
dispersion statistics using pairs of drifter trajectories and their
velocities in an attempt to deduce the shape of the kinetic energy
wavenumber spectrum \cite{LaCasce-Ohlmann-03, LaCasce-10, Poje-etal-14,
Beron-LaCasce-16, Zavala-etal-17a}.  Other work employed drifter
trajectory data to assess the significance of transport patterns
detected from altimetry-derived velocity using nonlinear dynamics
tools \cite{Olascoaga-etal-13, Beron-etal-15, Romero-etal-16}.
Additional work was more concerned with making practical use of the
drifter data through assimilating drifter velocities into ocean
general circulation models \cite{Coelho-etal-15} and blending these
velocities with altimetry-derived velocities to improve near-real-time
synoptic estimates of ocean currents \cite{Berta-etal-15}.  Descriptive
studies were also reported highlighting preferred synoptic pathway
patterns \cite{Yang-etal-99, DiMarco-etal-05, Perez-etal-13,
Zavala-etal-17b}.

The number of drifters that have surveyed the GoM is large enough
that a global characterization of the GoM's Lagrangian dynamics can
be sought.  This is indeed possible thanks to probabilistic tools
which enable sketching absorbing and almost-invariant sets in the
phase space of a nonlinear dynamical system \cite{Hsu-87,
Dellnitz-Junge-99, Froyland-05}.  The relevant nonlinear dynamics
here are the Lagrangian dynamics, which can be appropriately
discretized using those tools into a Markov chain described by a
matrix of transitional probabilities of moving between states of
the chain.  Inspection of the eigenvectors of the transition matrix
\cite{Froyland-etal-14} enables localization of regions of the flow
where trajectories converge in forward time as well as the regions
where those trajectories originate from (i.e., their backward-time
basins of attraction), thereby determining the connectivity between
separated locations in the flow domain.  This is conceptually very
different than the traditional approach to population connectivity
in marine systems, wherein transition matrices are constructed based
on ad-hoc partitions of the flow domain into putative spawning and
recruitment areas \cite{Cowen-etal-07}.  The eigenvector method of
Froyland et al.\ \cite{Froyland-etal-14} we employ here also differs
from the recent flow network approach \cite{Rosi-etal-14,
Sergiacomi-etal-15} as the former analyzes time-asymptotic aspects
of the dynamics through spectral information from the generating
Markov chain, while the latter computes various graph-based quantities
for finite-time durations to study flow dynamics. We note that
Froyland's et al.\ \cite{Froyland-etal-14} approach is also subtly
different from the eigenvector method of Froyland et al.\
\cite{Froyland-etal-07}, where structures that are approximately
fixed over a finite time duration were extracted.  Although attracting
regions may be small and trap trajectories for long periods of time
before eventually exiting and thus constituting almost-invariant
regions, if their basins of attraction are large, they can exert
great influence on the global Lagrangian dynamics.  Decompositions
of the surface-ocean flow into almost-invariant sets form the basis
of a \emph{Lagrangian dynamical geography}, where the boundaries
between basins are determined by the Lagrangian circulation itself,
instead of arbitrary geographical divisions.

Our goal in this paper is to construct such a Lagrangian dynamical
geography for the GoM, which carries useful information for guiding
activities dealing with hampering and/or palliating the effects of
an oil spill or a harmful algal bloom or supporting stock assessment
efforts and management decisions for fishing regulations, just to
cite some examples.

\section*{Results}

\begin{figure*}[t]
  \centering \includegraphics[width=\textwidth]{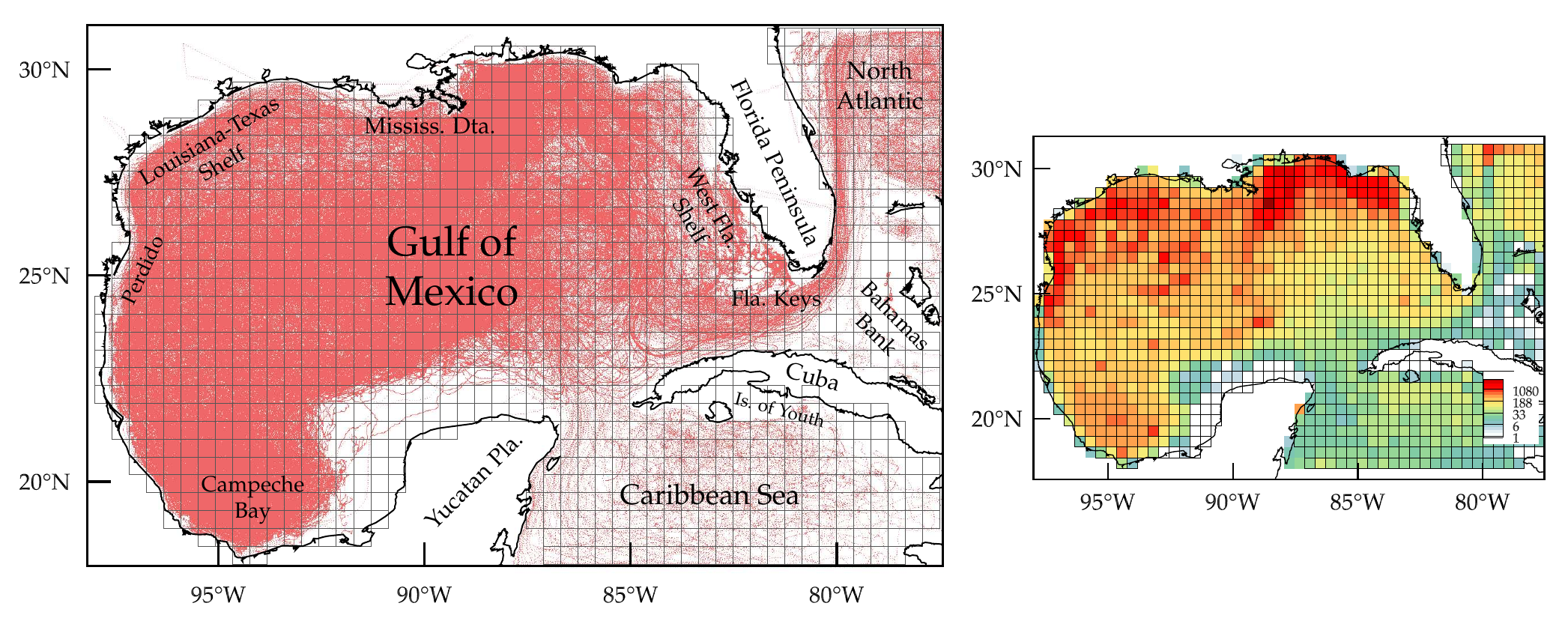}
  \caption{(left) ``Spaghetti'' plot of all daily satellite-tracked
  drifter trajectories describing the surface-ocean Lagrangian
  dynamics in the extended Gulf of Mexico (eGoM) domain.  The finite
  grid used to construct a Markov-chain representation of the
  dynamics is shown in black.  (right) Number of drifters per grid
  bin independent of the day over 1994--2016 subjected to a fourth-root
  transformation. Figures constructed using Matlab R2017a
  (http://www.mathworks.com/) and Tecplot 360 2016 R2
  (http://www.tecplot.com/).}
  \label{fig:bin}%
\end{figure*}

The left panel of Fig.\ \ref{fig:bin} shows all 3207 daily drifter
trajectories (in red) available to us over 1994--2016 for describing
the surface-ocean Lagrangian dynamics in the domain of interest.
The domain, referred to herein as extended GoM (eGoM), includes the
GoM itself and small adjacent portions of the Caribbean Sea and
North Atlantic.  The ``spaghetti'' plot shown in the figure reveals
that the drifters sample most of the domain (about 3 drifters are
found per km$^2$ on average ignoring time).  Exceptions are relatively
small regions on the Yucatan Shelf, south of Cuba, and the Bahamas
Bank, which have never been visited by any drifters.  As described
in some depth in Materials, the drifters differ in design from
experiment to experiment, so some variations in their Lagrangian
properties can be expected \cite{Beron-etal-16}.  For the purposes
of this work, these variations were ignored.

\paragraph{Construction of the Markov-chain model.}

In order to apply the eigenvector method, outlined in Methods, we
proceeded to discretize the Lagrangian dynamics as represented by
the drifter motion as follows.  We first laid down on the eGoM a
grid of $N = 1500$ square bins of (roughly) 50-km side as shown (in
black) in the left panel of Fig.\ \ref{fig:bin}.  While we did not
perform optimizations of any kind, this grid resolution resulted
in a reasonable individual bin coverage as shown in the drifter
density plot in the right panel of Fig.\ \ref{fig:bin}.  Ignoring
time, there are 246 drifters on average per bin, with as many as
4266 drifters in some bins and a few (108) empty bins.

We then constructed, according to \eqref{eq:P}, the $N \times N$
matrix $P$ of transitional probabilities of the drifters to move
between bins of the grid (i.e., states of the Markov chain defined
by $P$) using 2-day-long trajectory pieces, which may begin on any
day.  Neglecting the start day corresponds to assuming that the
Lagrangian dynamics are statistically stationary.  This is clearly
a strong assumption, which is commonly made in relative dispersion
analyses in connection with turbulence phenomenology studies
\cite{Bennett-06}.  This assumption may be more easily justified
when the focus is on long-time behavior , as is the case of this
work and earlier related works \cite{Maximenko-etal-12, vanSebille-etal-12,
Froyland-etal-14}.  On the other hand, 2-day-long trajectory pieces
are long enough to enable interbin communication in most cases.
Furthermore, 2 days is larger than the typical Lagrangian decorrelation
timescale near the ocean surface, estimated to be of about 1 day
\cite{LaCasce-08}. Thus there is negligible memory farther than 2
days into the past, and so the Markov assumption is approximately
satisfied.

An analysis of the sensitivity to transition time changes and data
reductions is presented in Appendix A of the Supplementary Information.
This reveals that the results discussed below do not depend on the
above specific choices.  Independent of these choices, a few bins
were always found to be initially empty, which had to be excluded,
making $P$ always slightly deviate from being row-stochastic (some
of its rows did not add up to 1 exactly).  This required us to
row-stochasticize $P$ by normalizing each row by its sum, following
standard procedures \cite{Froyland-etal-14}.

\paragraph{Communication within the Markov chain.}

\begin{figure*}[t]
  \centering
  \includegraphics[width=.5\textwidth]{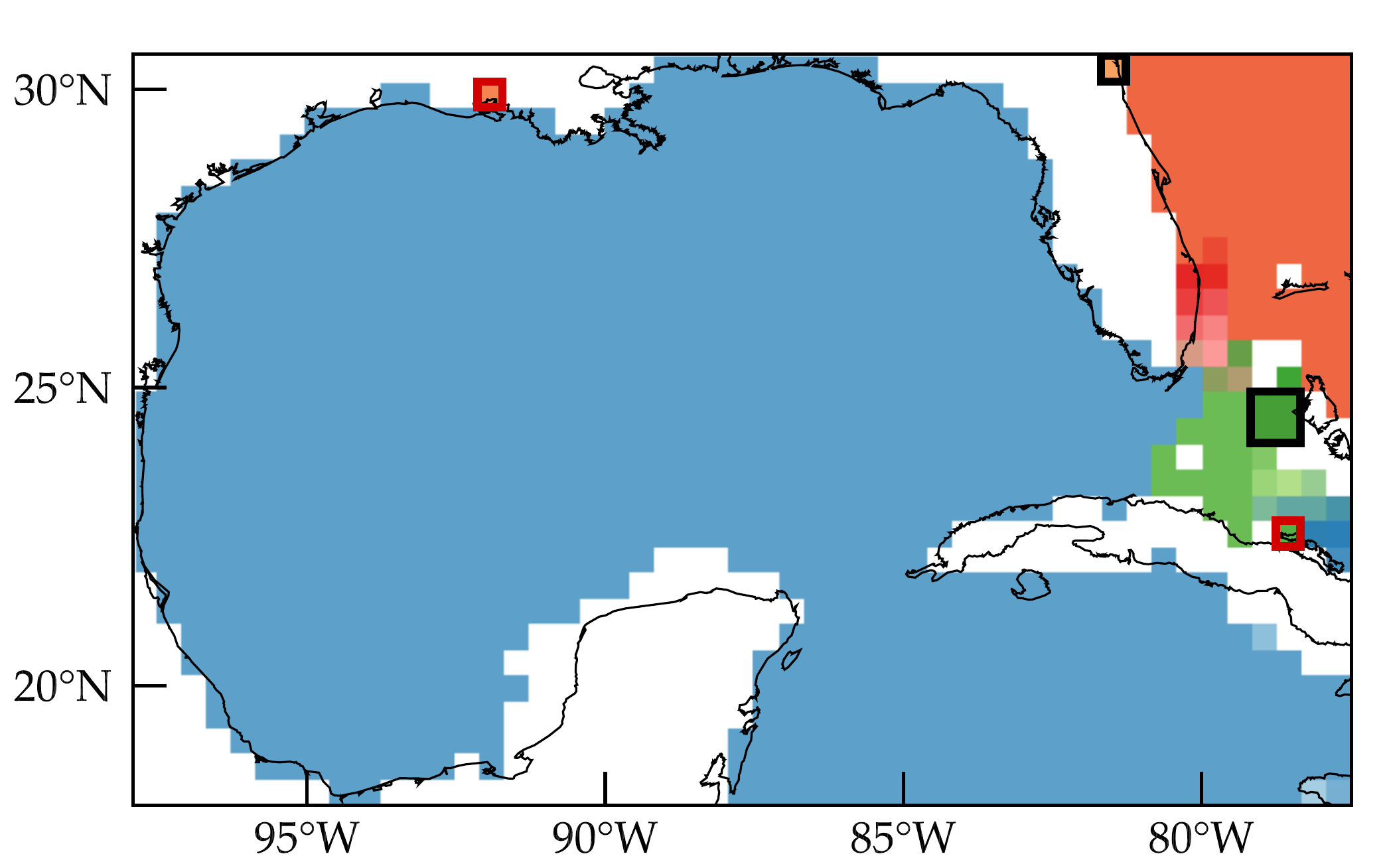}%
  \caption{Grouping of the Markov-chain states into classes according
  to their communication type. Bins indicated with the same color
  belong to the same class.  The are 31 communicating classes
  covering the eGoM domain.  Two of these classes are closed (red
  boxes) and 2 are both closed and absorbing (black boxes); they
  do not offer any insight into the Lagrangian dynamics other than
  mere beaching of drifters.  Computations carried using Matlab
  R2017a (http://www.mathworks.com/) and visualization using Tecplot
  360 2016 R2 (http://www.tecplot.com/).}
  \label{fig:comm}%
\end{figure*}

Once the transition matrix $P$ was constructed, we moved on to
determine the level of communication among the states of the Markov
chain. Let $S \subset \{1,\cdots,N\}$ represent a class of states;
the following communication types can be distinguished \cite{Norris-98}.
The class $S$ is said to be communicating if for each pair $i,j \in
S$ there exists $m > 0$ finite such that $(P^m)_{ij} > 0$ (i.e.,
there is a positive probability of moving between any pairs of
states in the class in a finite number of steps).  A communicating
class $S$ is said to be closed if it has $P_{ij} = 0$ for all $i
\in S$, $j \notin S$ (i.e., there is a zero probability of moving
out of the class to another state elsewhere in the computational
domain).  A communicating class $S$ is said to be absorbing if given
some $m > 0$ finite, $(P^m)_{ij} > 0$ for at least one $i \notin
S$, $j \in S$ (i.e., there is a positive probability of at least
one external state moving into the class in a finite number of
steps).  In practice a Markov chain can be viewed as a directed
graph with vertices in the graph corresponding to states in the
chain, and directed arcs in the graph corresponding to one-step
transitions of positive probability.  This allows one to apply
Tarjan's algorithm \cite{Tarjan-72} to assess communication within
a chain.  Specifically, the Tarjan algorithm takes such a graph as
input and produces a partition of the graph's vertices into the
graph's strongly connected components.  A directed graph is strongly
connected if there is a path between all pairs of vertices.  A
strongly connected component of a directed graph is a maximal
strongly connected subgraph and by definition also a maximal
communicating class of the underlying Markov chain.  Applying
Tarjan's algorithm to the Markov chain derived using the drifter
trajectory data we found a total of 31 strongly connected components
or, equivalently, maximal communicating classes.  Each one of these
classes is indicated with a different color in Fig.\ \ref{fig:comm}.
Direct verification further showed that, out of the 31 communication
classes revealed by the Tarjan algorithm, there are 2 closed
communicating classes, one consisting of a single state and another
one made up of four states; the two closed classes are highlighted
with red boxes in Fig.\ \ref{fig:comm}.  Two additional single-state
communicating classes (highlighted with black boxes in Fig.\
\ref{fig:comm}) were determined by direct verification to be both
closed and absorbing.

The closed communicating classes correspond to coastal or near-coastal
regions where the drifters initially lying inside do not move or
stay within, or initially lying outside get trapped within.  Other
than reflecting beaching, these classes do not represent any
dynamically relevant process.  Relevant from a Lagrangian dynamics
viewpoint is the large group of communicating classes, which cover
almost entirely the eGoM.  This group is composed of 2 large classes,
one covering the entire GoM and the Caribbean Sea portion of the
eGoM (indicated in blue) and the other one spanning a large part
of the North Atlantic sector (in orange), and 25 additional smaller
classes occupying the rest of the North Atlantic sector (in red and
darkgreen tones).  Communication among the states of this group is
permitted.  Communication with states outside of the group is
possible too due to the presence of the small absorbing classes.

The presence of the small closed communicating classes implies that
the forward evolution of initially arbitrary distributed tracers
on the eGoM under the action of the transition matrix $P$ will never
attain a unique invariant distribution.  A tracer initially covering
the entire eGoM will nevertheless be supported on these small classes
in the long run.  The tracer initially inside the regions occupied
by the closed classes will remain within at all times, while the
tracer outside of these regions will converge, eventually after a
very long time, into the regions occupied the absorbing classes.
But the dynamically relevant phenomena are transient, occur during
earlier times, and are not influenced by the small closed communicating
classes, which we have verified by direct computation.

\paragraph{Forward evolution of a tracer density.}
 
\begin{figure*}[t]
  \centering
  \includegraphics[width=\textwidth]{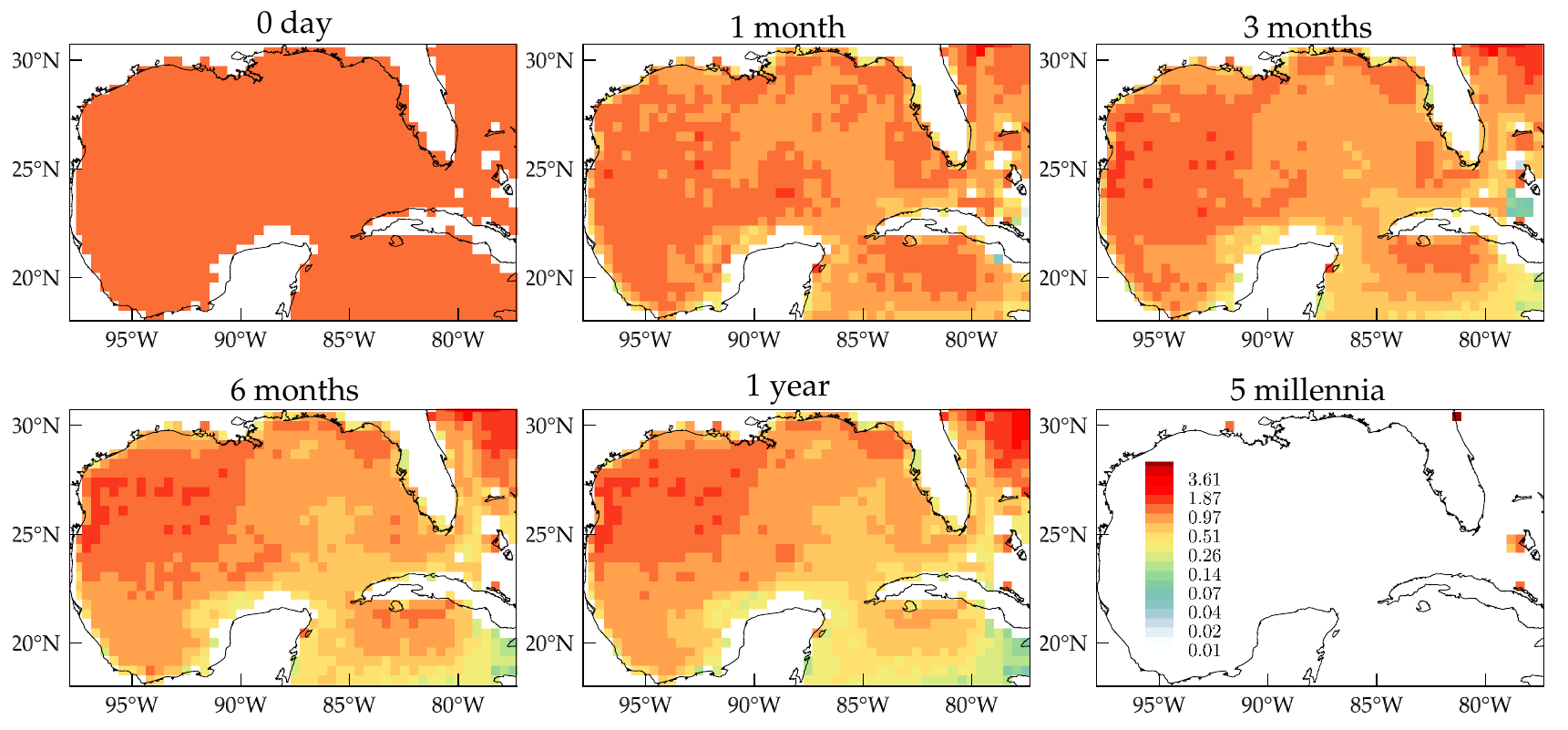}%
  \caption{Forward evolution of an initially uniformly distributed
  tracer under the action of the transition matrix $P$. The tracer
  concentration is normalized by the mean concentration and subjected
  to a fourth-root transformation. Computations carried using Matlab
  R2017a (http://www.mathworks.com/) and visualization using Tecplot
  360 2016 R2 (http://www.tecplot.com/).}
  \label{fig:push}%
\end{figure*}

Figure \ref{fig:push} shows selected snapshots of the forward
evolution (of the probability density) of a tracer, initially
uniformly distributed over the entire eGoM grid, under the action
of $P$, according to \eqref{eq:pP}.  It is immaterial to the method
whether $P$ models purely advective dynamics or a combination of
advective and diffusive dynamics \cite{Froyland-01}.  Because $P$
was constructed using 2-day-long drifter trajectory pieces, pushing
forward the tracer 1 step is equivalent to evolving the tracer in
forward time for 2 days.  As time advances the tracer distribution
loses homogeneity, increasing its density on various regions well
separated from one another on the northwestern and eastern sides
of the GoM, a region south of Cuba in the Caribbean Sea, and the
northeastern side of the North Atlantic portion of the eGoM.  These
regions show accumulation trends with varied levels of persistence,
thereby representing almost limiting invariant distribution regions.
Out of the four regions, the region lying in the North Atlantic
shows a sustained accumulation trend over several years (15 years
or so).  This accumulation, however, does not represent any real
aspect of the Lagrangian dynamics.  Rather, it reflects that the
eGoM is closed and the tracer must eventually exit the GoM and
Caribbean Sea sectors of the eGoM through the Straits of Florida
into the North Atlantic.  Considering that the GoM is fully drained
when the tracer mass within has decreased
by 95\%, we estimate a residence timescale for the GoM of about
13 years.  Accumulation on the other regions is much more meaningful
from a Lagrangian dynamics stand point.  Lasting for about 4 years,
accumulation in the northwestern side of the GoM on the southern
end of the Texas--Louisiana Shelf, more specifically a region lying
near the Perdido Foldbelt,  implies the existence of a persistent
convergent circulation along the coast from the south and north and
possibly also a mean westward flow.   The accumulation trend on the
northern part of the Caribbean Sea south of Cuba is shorter, of a
few months, possibly reflecting trapping inside transient closed
circulations there or accumulation induced by Ekamn transport
associated with trade winds.  The eastern side of the GoM over the
northern West Florida Shelf and the Florida Keys show a somewhat
less persistent accumulation trend, consistent with a relatively
rapid flow along the shelfbreak and out of the GoM through the
Straits of Florida.  As anticipated, the small closed communicating
classes do not determine the above accumulation regions or influence
their duration trends as it takes very long time (nearly 5 millennia!)
for the initially uniform distribution to settle on them.

\paragraph{Analysis of the Markov chain's eigenspectum.}

We can now turn to the application of the eigenvector method, which
enables a connection between structure in the eigenvector fields
of the transition matrix $P$ with (almost) invariant flow regions
(or sets) that attract tracer into as well as their corresponding
basins of attraction.

First we note that eigenvalue $\lambda = 1$ of $P$ is the largest
and further has multiplicity 4, consistent with the Markov chain
having 4 closed communicating classes of states.   The corresponding
left eigenvectors of $P$ reveal invariant distributions supported
on the small regions occupied by them (i.e., they do not change
under the action of $P$) and the right eigenvectors their basins
of attraction.  The invariant distributions are also limiting
distributions, as the forward evolution of a uniform tracer has
revealed.  As we have noted, they do not represent any relevant
aspect of the Lagrangian dynamics other than mere beaching.  Thus
we do not consider them any further here, but offer a discussion
for completeness in Appendix B of the Supplementary Information.

\begin{figure*}[t]
  \centering
  \includegraphics[width=.9\textwidth]{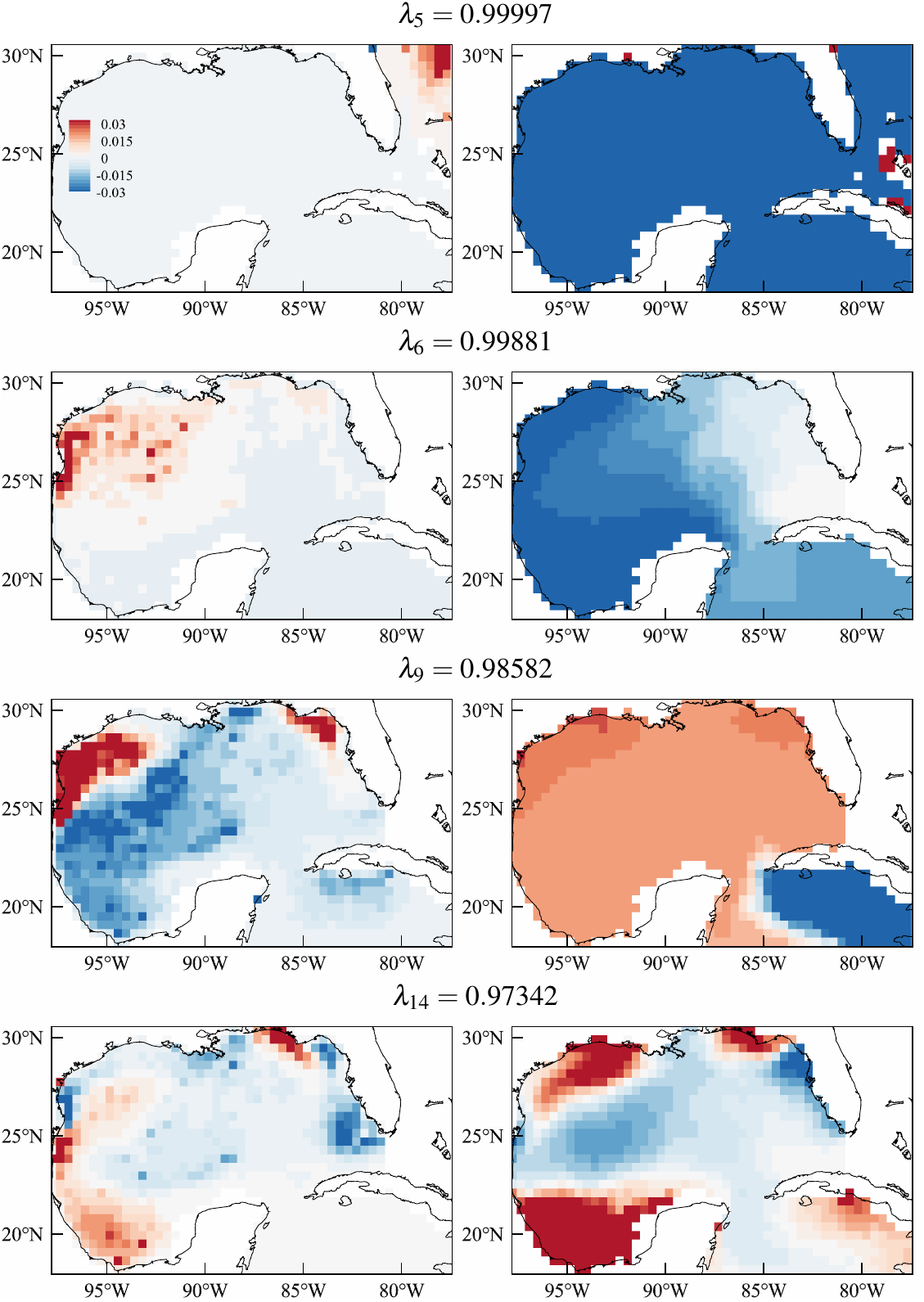}%
  \caption{(left) Selected left eigenvector fields of the transition
  matrix $P$ showing the locations of almost-invariant regions of
  forward-time attraction.  (right) Corresponding right eigenvector
  fields showing the backward-time basins of attraction for the
  attractors on the left. Computations carried using Matlab R2017a
  (http://www.mathworks.com/) and visualization using Tecplot 360
  2016 R2 (http://www.tecplot.com/).}
  \label{fig:evec}%
\end{figure*}

Dynamically relevant are the larger regions showing shorter but
still persistent accumulation trends as the initially uniformly
distributed tracer is pushed forward by $P$.  These can be revealed
by the eigenvector method as they have imprints on the left
eigenvectors of $P$ with $\lambda \approx 1$.  Their basins of
attraction have their footprints on the corresponding right
eigenvectors.  Figure \ref{fig:evec} shows 4 selected $\lambda
\approx 1$ left eigenvector fields on the left with the corresponding
right eigenvectors on the right.  The dominant ($\lambda = 0.99997$)
eigenvectors are shown in the top row.  The left eigenvector is
supported on the accumulation region on the northeastern corner of
the eGoM revealed from direct tracer forward evolution, thereby
revealing this region as an attracting set which remains almost-invariant
under the action of $P$.  As noted above this accumulation
reflects the fact that the GoM and Caribbean Sea portions of the
eGoM must be evacuated by tracers through the Straits of Florida.
Consistent with this, the right eigenvector identifies the basin
of attraction of this almost-invariant attracting set with the whole
eGoM modulo the small closed communicating class regions.  There
the right eigenvector is positive and indistinguishably flat.

Before proceeding to the analysis of the other eigenvectors, a few
remarks are in order.  First, the above large basin of attraction
may, and certainly does as we show below, include smaller basins
of attraction for other almost-invariant attracting sets contained
within.  Second,  those domains of attraction are weakly dynamically
interacting in the sense that they do not have perfect but rather
permeable boundaries on the long run, consistent with the fact that
$\lambda$ is not exactly unity.  Third, because eigenvectors with
nearly the same eigenvalue ($\lambda \approx 1$) can be linearly
combined to form an eigenvector with nearly the same eigenvalue,
the smaller almost-invariant attracting sets can also approximately
span a larger almost-invariant attracting set through linear
combination.

With the above observations in mind, we now proceed to analyze the
remaining eigenvectors. The second row of Fig.\ \ref{fig:evec} shows
left and right eigenvectors with $\lambda = 0.99881$, the second
largest nonunity eigenvalue.  The left eigenvector (left panel) is
supported on a region on the northwestern GoM covering mainly the
southern part of the Texas--Louisiana Shelf near the Perdido Foldbelt,
which coincides with the area revealed from direct tracer evolution
that showed the second longest accumulation trend.  This region is
now identified as an almost-invariant attracting set by the eigenvector
method.  The right eigenvector (right panel) reveals a partition
of the GoM into two halves by a nearly straight boundary roughly
connecting the Mississippi River Delta and the easternmost tip of
the Yucatan Peninsula.  On the western half the right eigenvector
takes a nearly constant positive value, identifying this region
with the basin of attraction for the almost-invariant attracting
set revealed by the left eigenvector.

The $\lambda = 0.98582$ left and right eigenvectors reveal additional
well-defined almost-invariant attracting sets and basins of attraction
(Fig.\ \ref{fig:evec}, third row).  Specifically, a large region
on the Texas--Louisiana Shelf, a portion of the northern West Florida
Shelf, a domain south of Cuba east of the Island of Youth, and an ample
region on the western side of the GoM are revealed as almost-invariant
areas of attraction (left panel).  The corresponding basins of
attraction (right panel) span larger regions including those areas.

The $\lambda = 0.97342$ left--right eigenvector pair unveils further
almost-invariant attracting sets and basins of attraction (Fig.\
\ref{fig:evec}, bottom row).  Most notably, the Florida keys are
identified as an almost-invariant region of attraction by the left
eigenvector and the West Florida Shelf as its basin of attraction
by the right eigenvector.  Likewise, the southwestern portion of
the Bay of Campeche is revealed as an almost-invariant attracting
set with the corresponding basin of attraction spanning the whole
Bay of Campeche.

\begin{figure*}[t]
  \centering
  \includegraphics[width=.75\textwidth]{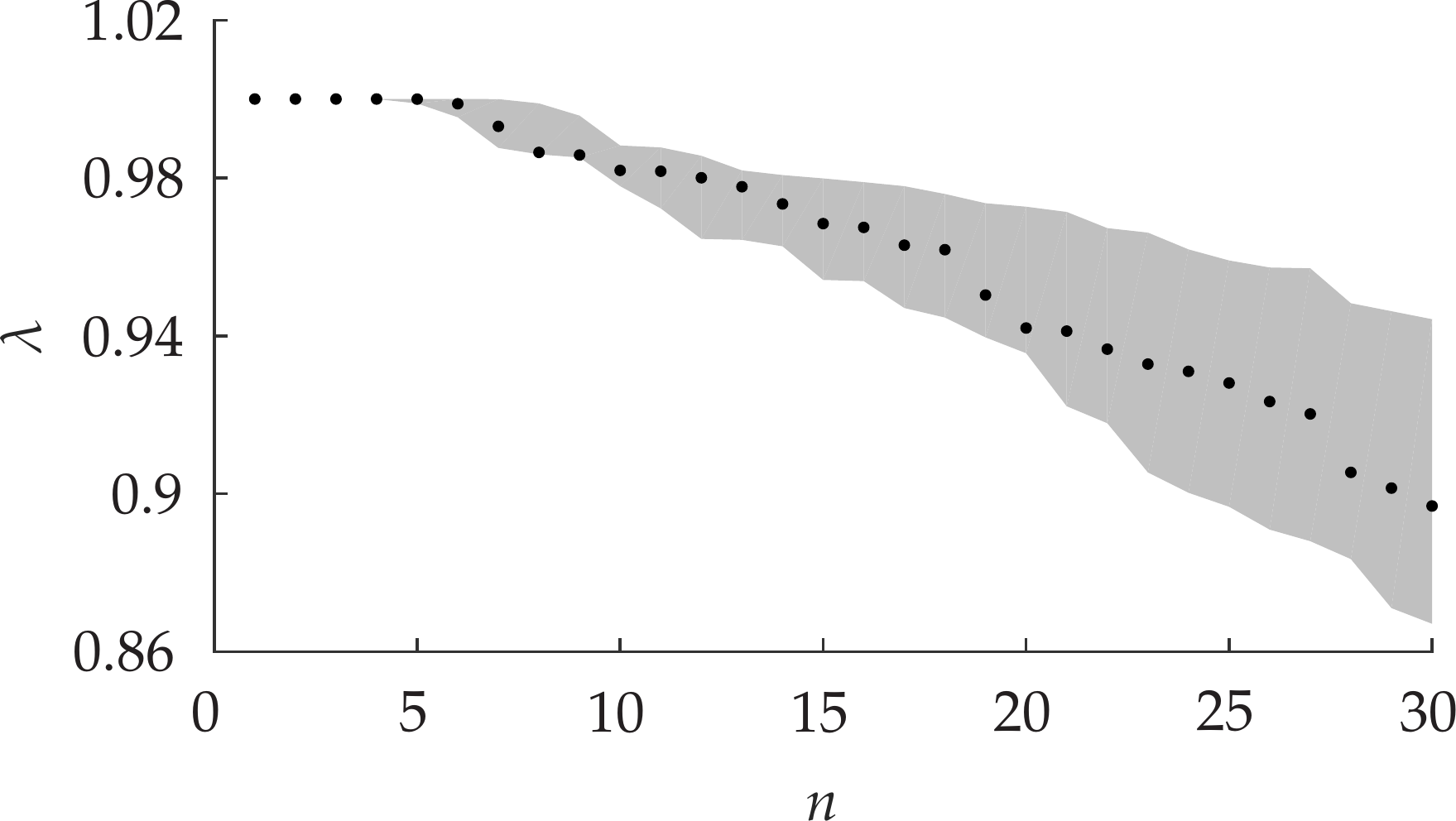}%
  \caption{First 30 eigenvalues of the transition matrix $P$ computed
  using two-day-long drifter trajectory data (dots) and uncertainties
  (gray shade) representing the total variation across eigenvalues
  computed from an ensemble of $P$s produced by randomly perturbing
  the trajectories.  Computations and visualization carried using
  Matlab R2017a (http://www.mathworks.com/).}
  \label{fig:error}%
\end{figure*}

Left--right eigenvector pairs with $\lambda > 0.97342$ not discussed
above do not provide any additional relevant information.  Indeed,
almost-invariant attractors and basins of attraction covering
locations similar to those described above are in general revealed
(a subset of these eigenvectors are shown in Fig.\ S4 of the
Supplementary Information).  Left--right eigenvector pairs with
$\lambda < 0.97342$ can be ignored because of their lower relative
statistical significance.  Figure \ref{fig:error} shows a portion
of the discrete eigenspectrum of $P$ and an estimate of the associated
uncertainty.  Specifically, the dots correspond to the first 30
eigenvalues (out of a total of $N = 1500$).  For the $n$th eigenvalue
the width of the grey shading corresponds to the maximum variation
across an ensemble of 100 realizations of the $n$th eigenvalue
computed from transition matrices constructed using randomly perturbed
drifter trajectories with an amplitude ranging from  500 m to 1.5
km, i.e., from at least as 100 times as large as the Global Positioning
System (GPS) accuracy to up 10 times large as the \emph{Argos}
satellite system mean positioning error.  Note that the uncertainty
grows most noticeably starting at around the 15th eigenvalue,
$\lambda \approx 0.96841$, suggesting a cutoff for eigenvector
analysis.

A timescale $\tau$ characterizing the level of invariance of a set
can be calculated by thinking of $\lambda < 1$ as a decay rate.
Let $T$ denote the transition time between states of the Markov
chain represented by the transition matrix $P$.  Let $\alpha < 1$
be the fraction by which a set has decayed under $n$ applications
of $P$.  Then $\tau = \frac{\log\alpha}{\log\lambda}T$.  Taking
$\alpha = 0.5$, representing a set's ``half life,'' the invariance
timescales of the attracting sets identified in the left column of
Fig.\ \ref{fig:evec} approximately are, from top to bottom, $\tau
= $ 13 years, 3 years, 3 months, and 2 months.  These invariance
timescales are consistent with the accumulation persistence times
roughly estimated above from the forward evolution of a uniform
tracer distribution.

\paragraph{Construction of the Lagrangian dynamical geography.}

\begin{figure}[t]
  \centering
  \includegraphics[width=\textwidth]{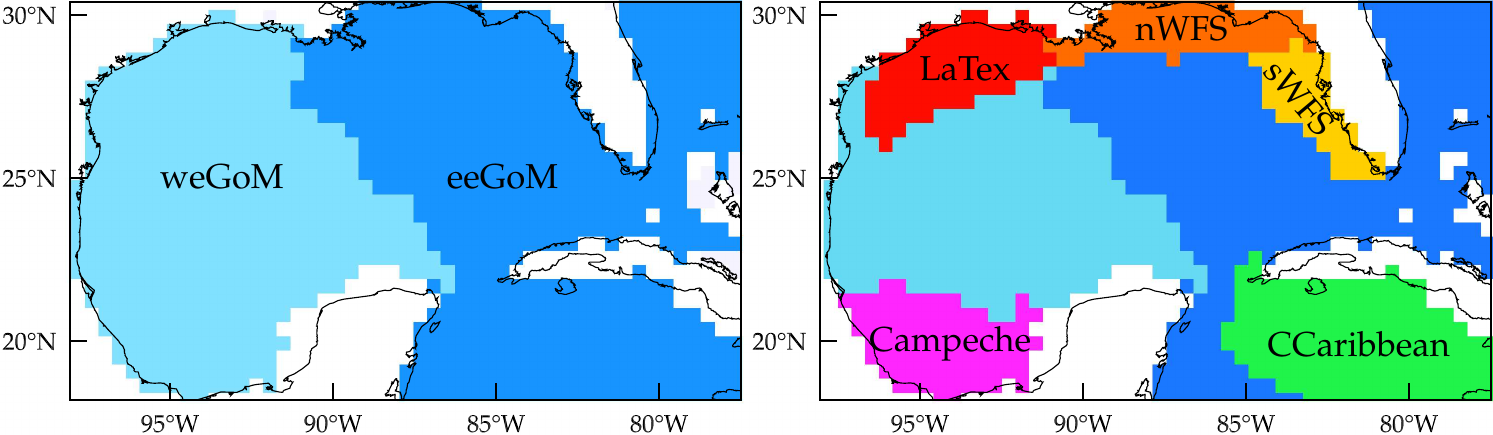}%
  \caption{Main (left) and refined (right) dynamical geographies
  with weakly interacting provinces corresponding to domains of
  attraction associated to the largest almost-invariant attractors
  identified. Computations carried using Matlab R2017a
  (http://www.mathworks.com/) and visualization using Tecplot 360
  2016 R2 (http://www.tecplot.com/).}
  \label{fig:geo}%
\end{figure}

With the knowledge acquired using the eigenvector method we can now
move on constructing a Lagrangian dynamical geography for the eGoM.
This is done by patching together the weakly communicating basins
of attraction revealed through an appropriate thresholding of the
right eigenvectors.  We have found that setting the threshold at
0.005 gives satisfactory results; similar thresholds have been
considered in earlier work \cite{Froyland-etal-14}. The trivial
partition has a single province covering virtually the whole eGoM.
The coarsest nontrivial partition (Fig.\ \ref{fig:geo}, left panel)
has two dynamical provinces separated by a nearly straight boundary
connecting the Mississippi River Delta and the easternmost tip of
the Yucatan Peninsula.   The eastern eGoM province is labeled eeGoM,
while the western eGoM province is denoted weGoM.  A refined partition
(Fig.\ \ref{fig:geo}, right panel) has five additional dynamical
provinces roughly spanning the northern West Florida Shelf, the
southern West Florida Shelf, the Louisiana--Texas Shelf, the Bay
of Campeche, and a large region over the Cuban Caribbean Sea.  These
provinces are denoted nWFS, sWFS, LaTex, Campeche, and CCaribbean,
respectively.  Tracers initially within these provinces will spend
more time moving and eventually temporarily accumulating in regions
within them than dispersing across their boundaries.  These dynamical
provinces thereby set the way that distant regions in the eGoM, and
the GoM in particular, are connected by tracer motion.

\begin{figure}[t]
  \centering
  \includegraphics[width=.9\textwidth]{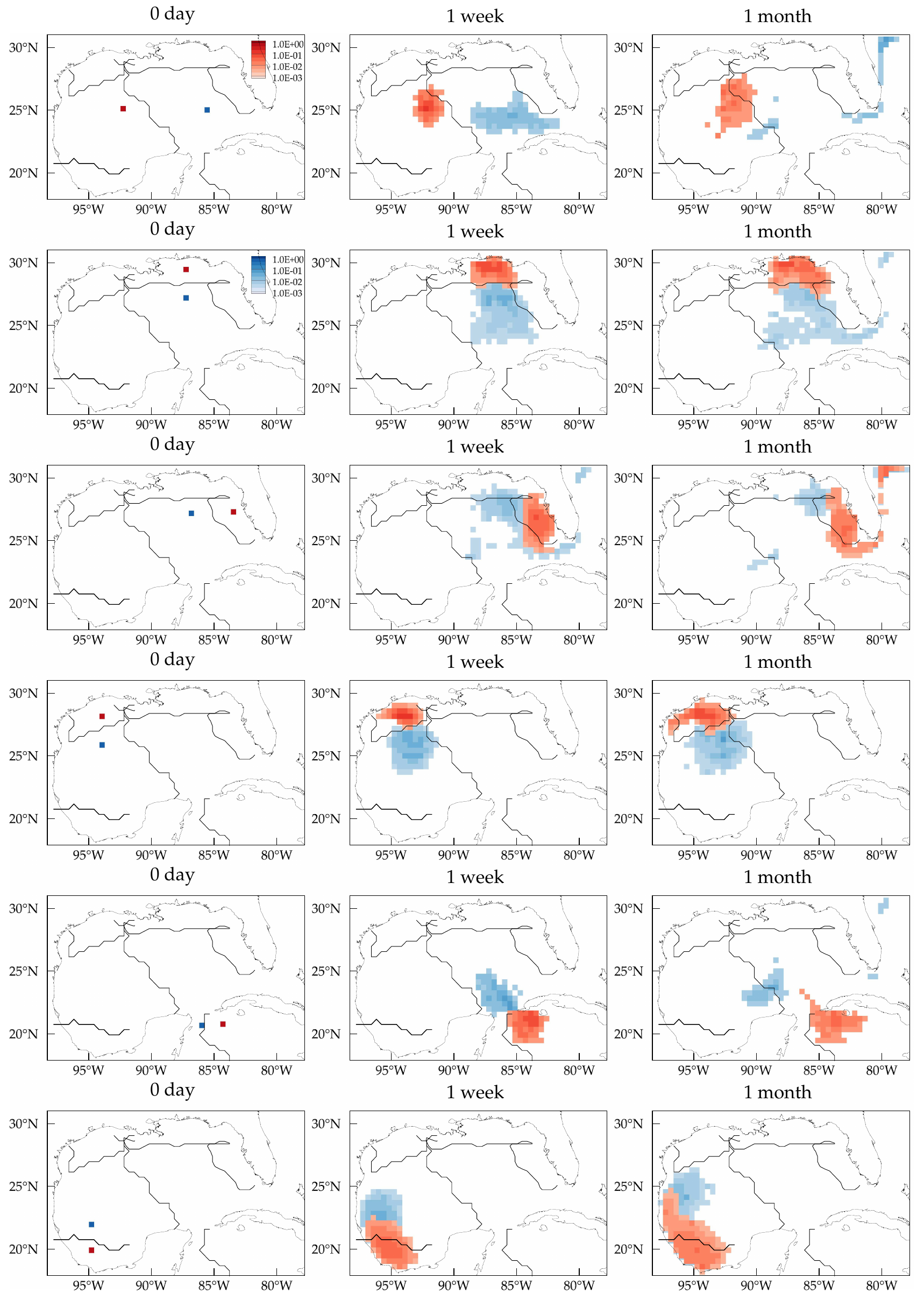}%
  \caption{Forward evolution under the action of the transition
  matrix $P$ of tracers released at various localized regions with
  dynamical geographic divisions overlaid.  The tracer concentration
  is normalized by the mean concentration. Computations carried
  using Matlab R2017a (http://www.mathworks.com/) and visualization
  using Tecplot 360 2016 R2 (http://www.tecplot.com/).}
  \label{fig:source}%
\end{figure}

To illustrate the impact of the Lagrangian dynamical geography on
the Lagrangian circulation explicitly, forward tracer evolutions
from various point sources are shown in Fig.\ \ref{fig:source}.
The source locations were chosen to straddle some of the dynamical
province boundaries.  Note that while tracers are released relatively
nearby, they take on quite different paths.   Note also that the
level of leakage through the boundary of a dynamical province is
influenced locally by the level of invariance of the attracting
region contained within the province and remotely by that of any
attractors outside of the province but sufficiently close to it.
For instance, leakage through the border of the Campeche and LaTex
provinces is larger than through the border between the weGoM and
eeGoM provinces.  This is consistent with the attracting sets
contained inside the Campeche and LaTex provinces having a shorter
invariance timescale (2 months) compared to the invariance timescale
of the attracting region lying near the Perdido Foldbelt (3 years),
which remotely affects the Lagrangian dynamics inside the Campeche
and LaTex provinces.

\begin{figure}[t]
  \centering
  \includegraphics[width=.9\textwidth]{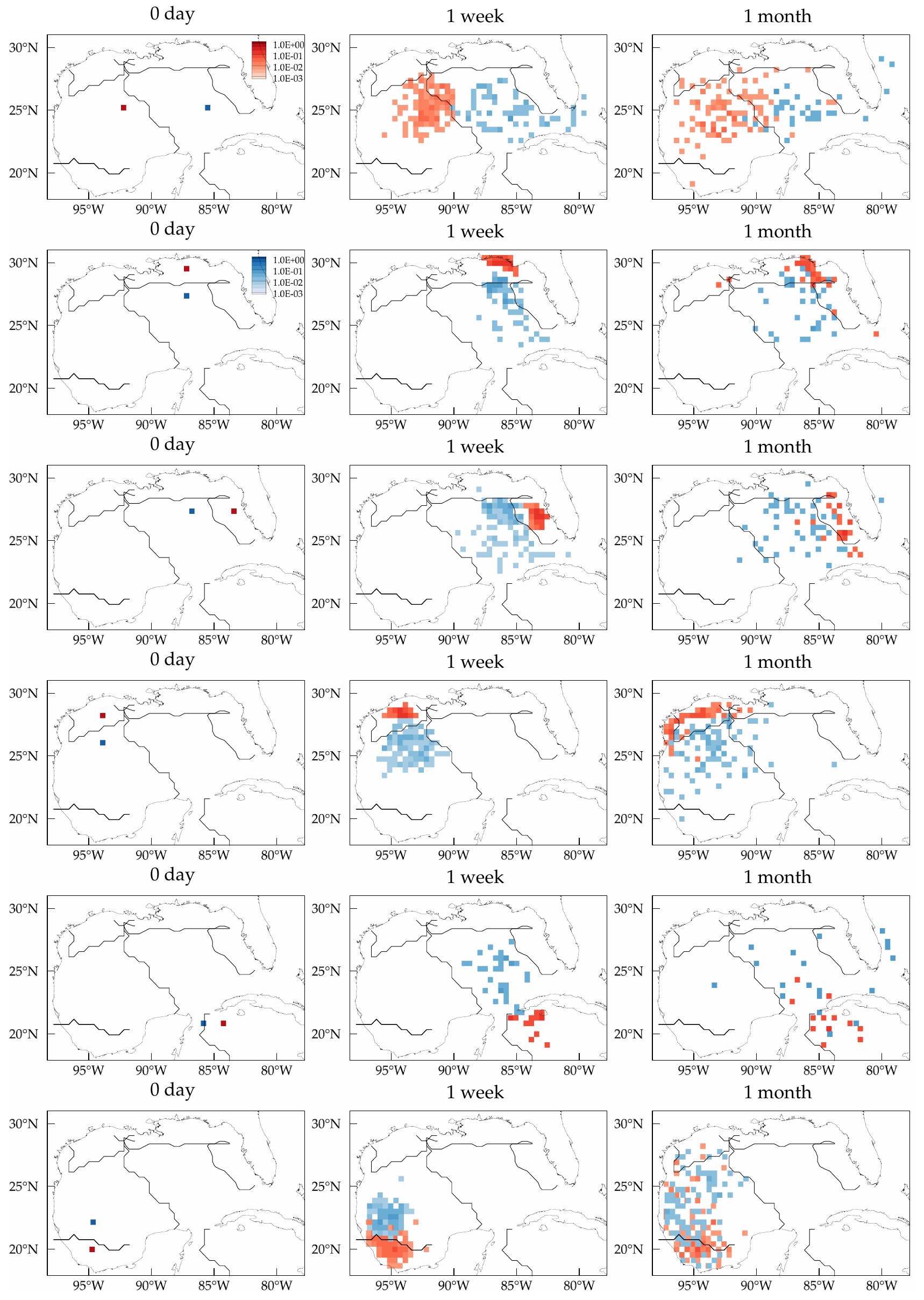}%
  \caption{Evolution of normalized drifter densities for all drifters
  that have gone through the locations indicated in the left column
  at some point in time.  Computations carried using Matlab R2017a
  (http://www.mathworks.com/) and visualization using Tecplot 360
  2016 R2 (http://www.tecplot.com/).}
  \label{fig:ensemble}%
\end{figure}

A verification of the significance of the Lagrangian dynamical
geography \emph{independent} of the Markov model derived here is
given in Fig.\ \ref{fig:ensemble}.  This figure shows evolutions
of positions of drifters, plotted as densities, for drifters going
through the same tracer source locations as in Fig.\ \ref{fig:source}.
More specifically, to create these densities, for all drifters that
have gone through these locations at some point in time, their
positions were recorded after 7 days and 1 month, and the number
of drifters falling in a given bin of of the grid in Fig.\ \ref{fig:bin}
counted and divided by the total number of drifters involved. Note
that the drifter densities evolve in a way that is broadly consistent
with the constraints imposed by the boundaries of the Lagrangian
dynamical geography.

\paragraph{Independent observational support for the geography.}

We close with an account of independent analyses of various different
observations that provide additional reality checks to the eigenvector
method results and thus the Lagrangian dynamical geography implied.
Analysis of satellite altimetry data \cite{Sturges-16} and ship-drift
\cite{DiMarco-etal-05} and drifter trajectory data \cite{Sturges-Kenyon-08}
with techniques different than those employed here have revealed
the existence of a mean westward flow in the GoM.  This provides a
mechanism for accumulation on western side of the GoM, and hence
independent sustain for the almost-invariant attracting set the
eigenvector method detected near the Perdido Foldbelt.  The persistent
cyclonic gyres south of Cuba observed in drifter trajectory data
\cite{Centurioni-Niiler-03} provide a mechanism for temporary
retention in the region, and hence support for the almost-invariant
attracting set identified there.  Satellite-tracked drifter
trajectories reveal persistent cyclonic motion in Bay of Campeche
\cite{Perez-etal-13, Zavala-etal-17b}, providing too a mechanism
for temporary retention consistent with our findings.  In turn,
Yang et al.\ \cite{Yang-etal-99} noted that satellite-tracked
drifters deployed on the northern GoM tend avoid the southern West
Florida Shelf, calling this region a ``forbidden zone.'' Such a
forbidden zone, whose boundary was later characterized using nonlinear
dynamics techniques \cite{Olascoaga-etal-06, Olascoaga-NPG-10},
forms one of the provinces of our dynamical geography.  Analyzing
card and bottle landings over 1955--1987 as well as marine mammals
and turtle strandings, Lugo-Fern\'andez et al.\ \cite{Lugo-etal-01}
reported coastal areas acting as attractors in near coincidental
position with some of those identified by the left eigenvectors of
the transition matrix constructed here.  Furthermore,  these authors
added ``\ldots the surface currents from ship-drift suggest that
the Gulf [of Mexico] may be bisected by a line along 86$^{\circ}$
or 87$^{\circ}$W to the Mississippi Delta.'' This is consistent
with the minimal partition of the eGoM resulting from the application
of the eigenvector method.

\section*{Conclusions}

Using an unprecedentedly large collection of satellite-tracked
surface drifter trajectory data in the GoM and its vicinity, we
constructed a Markov-chain representation of the GoM's surface-ocean
Lagrangian dynamics.  Analyzing the eigenvectors of the transition
matrix of the chain we identified almost-invariant regions of
attraction and their basins of attraction.  With this information
we then constructed a Lagrangian dynamical geography with weakly
interacting provinces.  Constraining the connectivity between distant
places in the GoM, the dynamical geography constructed can have
implications for offshore oil exploration, oil spill contingency
planning, pollution mitigation, and fish larval connectivity
assessment among many other activities of practical interest.

Our results were found to be robust under arbitrary data reductions,
but biases due to irregular spatiotemporal sampling may still be
possible.  While some drifters in the database cover extended periods
of time, representing quite well all possible dynamical scenarios,
a large number of drifters were deployed in localized regions in
the northern GoM, sampling particular dynamical conditions observed
during the months over which the experiments lasted.  Other drifters
targeted particular dynamical features like Loop Current rings,
potentially biasing accumulation on the western GoM.  Currently
underway is an assessment of the importance of these potential
biases using synthetic drifter trajectories as generated by different
ocean general circulation models.

\section*{Methods}

\paragraph{Eigenvector method.}

The eigenvector method \cite{Froyland-etal-14} employed here is
rooted in Markov-chain theory concepts that have previously been
used to approximate invariant sets in dynamical systems using
short-run trajectories \cite{Hsu-87, Dellnitz-Junge-99, Froyland-05}.
The dynamical system of interest is that governing the motion of
fluid particles, which are described by satellite-tracked drifters
on the ocean surface.

Let $X$ be a closed flow domain on the plane and denote by $T(x)$
the end point of a trajectory starting at $x \in X$ after some short
time.  A discretization of the dynamics can then be attained using
Ulam's method \cite{Ulam-79, Froyland-01} by dividing the domain
$X$ into $N$ connected bins $\{B_1,\cdots,B_N\}$, which will here
be assumed of the same area for simplicity.  The proportion of mass
in $B_i$ mapped to $B_j$ under one application of $T$ is (approximately)
equal to
\begin{equation}
   P_{ij} = \frac{\#\text{ of particles in }B_i\text{ that
   are mapped to }B_j}{\#\text{ of particles in }B_i}.
  \label{eq:P}
\end{equation}
The \emph{transition matrix} $P$ defines a Markov-chain representation
of the dynamics, with the entries $P_{ij}$ equal to the conditional
transition probabilities between bins, representing the states of
the chain. Let $\mathbf{f} = (f_1\, \cdots \, f_N)$ be the discrete
representation of $f(x)$, the probability density of some tracer
distribution.  Its forward evolution is calculated under right
multiplication by $P$, i.e.,
\begin{equation}
  \mathbf{f}^{(k)} = \mathbf{f}P^{k},\quad k = 1, 2, \cdots.
  \label{eq:pP}
\end{equation}
The above is the discrete representation of the push forward of
$f(x)$ under the action of the flow map $T$, namely, $\mathcal{P}f(x)
= f \circ T^{-1}(x)\,|\det \mathrm{D}T^{-1}(x)|$, where $\mathcal{P}$
is called a \emph{transfer} or \emph{Perron--Frobenius operator}
\cite{Froyland-05}.  If the flow is incompressible, in which case
$\det \mathrm{D}T(x) = 1$, then $\mathcal{P}f(x)$ corresponds to
an area-preserving rearrangement of $f(x)$.

Assume that $P$ is \emph{regular}, i.e., there exists $N < \infty$
such that $(P^n)_{ij} > 0$ for all $n \ge N$ and all $1 \le i,j \le
N$.  This means that $P$ is both \emph{irreducible} (i.e., for each
$1 \le i,j \le N$, $(P^{n_{ij}})_{ij} > 0$ for some $n_{ij} <
\infty$, so all states communicate) and \emph{aperiodic} (i.e., for
each $1 \le i \le N$ there exists $M < \infty$ such that $(P^m)_{ii}
> 0$ for all $m \ge M$, so no state occurs recurrently). By the
Perron--Frobenius theorem, the eigenvalue spectrum of $P$ satisfies
$|\lambda| \leq 1$.  The eigenvalue $\lambda = 1$ is simple and the
corresponding left and right eigenvectors both are positive.
Furthermore, there exists a unique limiting distribution, given by
\begin{equation}
  f_i^{(\infty)} = (P^\infty)_{ij} > 0\,\forall j,
  \label{eqPinfty}
\end{equation}
which satisfies
\begin{equation}  
  \mathbf{f}^{(\infty)} = \mathbf{f}^{(\infty)}P.
  \label{eq:Pinv}
\end{equation}
Note that the limiting distribution is an invariant distribution
(it does not change under the action of $P$).  Moreover, it is a
left eigenvector of $P$ with eigenvalue $\lambda = 1$. The associated
right eigenvector is $\mathbf{1} = (1\, \cdots \, 1)^\top$, as
required by row-stochasticity of $P$, viz.,
\begin{equation}
	 P\mathbf{1} = \mathbf{1}.
	\label{eq1}
\end{equation}

When $P$ is not regular, no unique limiting distribution exists,
and a nontrivial connection between left and right eigenvectors of
$P$ with invariant attracting sets and basins of attraction,
respectively, can be established.

\begin{figure*}[t]
  \centering
  \includegraphics[width=.75\textwidth]{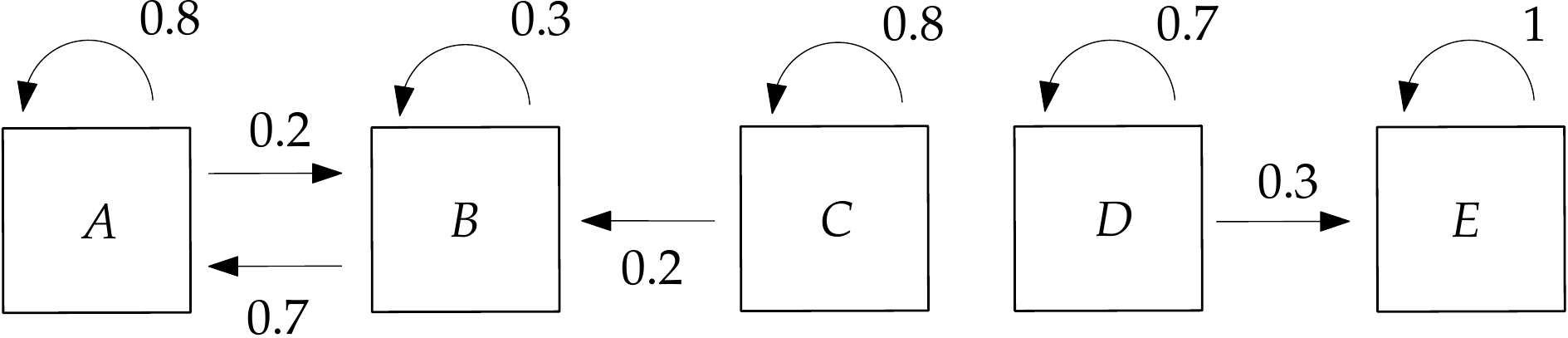}%
  \caption{Reduced Markov chain comprising 5 states with transition
  probabilities of moving among states indicated which is used to
  illustrate the idea of the eigenvector method. Figure generated
  using Ipe 7.2.7 (http://ipe.otfried.org/).}
  \label{fig:chain}%
\end{figure*}

To illustrate the above, consider the reduced chain comprising 5 states
$\{A,B,C,D,E\}$ represented by
\begin{equation}
    P = 
    \begin{blockarray}{lccccc}
         & A  & B & C & D & E\\
        \begin{block}{l(ccccc)}
			 A & 0.8 & 0.2 & 0   & 0   & 0  \\
          B & 0.3 & 0.7 & 0   & 0   & 0   \\
          C & 0   & 0.2 & 0.8 & 0   & 0   \\
		    D & 0   & 0   & 0   & 0.7 & 0.3 \\
			 E & 0   & 0   & 0   & 0   & 1   \\
        \end{block}
    \end{blockarray}
    \label{eq:P5}
\end{equation}
The chain with the transition probabilities among pairs of states
and corresponding directions indicated is sketched in Fig.\
\ref{fig:chain}.

States $\{A,B,C\}$ do not communicate with states $\{D,E\}$ and
thus the chain is not irreducible and therefore not regular.  In
particular, classes $\{A,B\}$ and $\{E\}$ are closed ($P = 0$ for
entries linking states in and outside of the class) and furthermore
absorbing ($P > 0$ for at least one entry linking a state outside
of the class with one in the inside thereof).  This implies the
existence of two limiting distributions, one supported on states
$\{A,B\}$ and another one supported on state $\{E\}$, which are
reached from any distributions initially supported on their basins
of attraction, given by states $\{A,B,C\}$ and $\{D,E\}$, respectively.
Direct computation gives $\mathbf{f}_{AB}^{(\infty)} =
(\frac{7}{9}\,\frac{2}{9}\,0\,0\,0)$ and $\mathbf{f}_{E}^{(\infty)}
= (0\,0\,0\,0\,1)$, respectively.  Linear combinations
$a\mathbf{f}_{AB}^{(\infty)} + b \mathbf{f}_{E}^{(\infty)}$, $a, b
\neq0$, also are limiting distributions, which are reached from
arbitrary distributions with support on $\{A,B,C\}$ and $\{D,E\}$.

Now, because the chain is not regular, the largest eigenvalue of
$P$, $\lambda = 1$, is not simple and in this case has multiplicity
2.  The left eigenvectors with $\lambda = 1$ are given by limiting
distributions $\mathbf{f}_{AB}^{(\infty)}$ and $\mathbf{f}_{E}^{(\infty)}$.
The corresponding right eigenvectors are given by $(1\,1\,1\,0\,0)^\top$
and $(0\,0\,0\,1\,1)^\top$, which are supported on the basins of
attraction of these limiting distributions; cf.\ Norris \cite{Norris-98},
thm.\ 1.2.3 or Froyland et al.\ \cite{Froyland-etal-14}, thm.\ 1.

The above illustrates the essence of the eigenvector method: invariant
attracting sets and disjoint basins of attraction can be revealed
from the structure of the left and right eigenvectors of $P$ with
eigenvalue 1, respectively.  This motivates the possibility of
revealing almost-invariant attracting sets and weakly disjoint
basins of attraction, relevant in dynamical systems governing motion
that exhibits transient behavior, by inspecting left and right
eigenvectors of $P$ with eigenvalues close to 1.

\paragraph{Data.} 

The analysis was performed using a large collection of drifter
trajectory data over 1994 through 2016. A total of 3207 drifter
trajectories from several different sources were considered.
Specifically, 1002 quarter-hourly trajectories of drifters deployed
during the LAgrangian Submesoscale ExpeRiment (LASER); 575 quarter-daily
trajectories from the National Oceanic and Atmospheric Administration
(NOAA) Global Drifter Program (GDP); 523 daily trajectories from
the Surface Current Lagrangian-Drift Program (SCULP); 441 hourly
trajectories from the Horizon Marine Inc.'s Eddy
Watch\textsuperscript{\textregistered} program; 372 trajectories
from the Centro de Investigaci\'on Cient\'{\i}fica y de Educaci\'on
Superior de Ensenada (CICESE)--Petr\'oleos Mexicanos (Pemex)
``Caracterizaci\'on Metoce\'anica del Golfo de M\'exico;'' 302
quarter-hourly trajectories from the Grand LAgrangian Experiment
(GLAD); 68 quarter-hourly from the NOAA/Atlantic Oceanic and
Meteorological Laboratory (AOML) South Floida Program (SFP); and
30 hourly drifters recorded by the U.S.\ Coast Guard (UCSC) during
LASER.

The LASER experiment used biodegradable drifters designed by
researchers of the Gulf of Mexico Research Initiative's Consortium
for Advanced Research on Transport of Hydrocarbon in the Environment
(CARTHE) at the University of Miami's Rosenstiel School of Marine
and Atmospheric Science.  They consisted of two main parts, a surface
wheel-shape float of 38-cm diameter and a cross-shape drogue made
of 38-cm $\times$ 38-cm rigid panels, both connected by a 15-cm
flexible tether. The positions of the drifters were tracked by GPS.
The GDP drifters \cite{Lumpkin-Pazos-07} follow the SVP (Surface
Velocity Program) design \cite{Sybrandy-Niiler-91}, consisting of
a spherical float which is drogued with a holey sock at 15 m.  Their
positions are tracked by the \emph{Argos} system or GPS.  The
drifters from SCULP \cite{Sturges-etal-01, Ohlmann-Niiler-05}
followed the Coastal Ocean Dynamics Experiment (CODE) design
\cite{Davis-85}.  These drifters consist of a 1-m-long rigid tube
with 4 sails attached forming a cross; floatation is supplied by
small floats fasten to the sails.  Their positions were tracked
using the \emph{Argos} system.  The drifters from the
EddyWatch\textsuperscript{\textregistered} program and the CICESE--Pemex
project are Far Horizon Drifters (FHD), manufactured by Horizon
Marine Inc.\ \cite{Sharma-etal-10, Anderson-Sharma-08}.  They consist
of a cylindrical buoy attached to a 45-m tether line, attached
itself to a 1.2-m ``para-drogue.'' These instruments are deployed
by air, so the drogue serves both as parachute to protect the buoy
when air deployed, as well as to reduce wind slippage of the buoy
as it drifts in the water. Positions are tracked using GPS. Most
drifters from GLAD \cite{Olascoaga-etal-13, Poje-etal-14,
Beron-LaCasce-16} were CODE type and GPS tracked.  Finally, the SPF
and USCG drifters are also CODE-type drifters with positions tracked
by the \emph{Argos} system.

For the purposes of the present work, which focuses on long-time
behavior,  it was sufficient to consider daily trajectories, which
required us to subsample those trajectories recorded at a higher
than daily rate.

\bibliography{fot}

\section*{Acknowledgment}

The constructive criticism of two anonymous reviewers led to
improvements in the manuscript.  The GLAD (doi:10.7266/\allowbreak
N7VD6WC8) and LASER (doi:110.7266/\allowbreak N7W0940J) drifter
trajectory datasets are publicly available through the Gulf of
Mexico Research Initiative Information and Data Cooperative (GRIIDC)
at https://data.\allowbreak gulfresearchinitiative.\allowbreak org.
The NOAA/GDP dataset is available at http://www.\allowbreak
aoml.noaa.gov/\allowbreak phod/dac. The drifter trajectory data
from Horizon Marine Inc.'s EddyWatch\textsuperscript{\textregistered}
program been obtained as a part of a data exchange agreement between
Horizon Marine Inc.\ and CICESE--Pemex.  The CICESE--Pemex
``Caracterizaci\'on Metoce\'anica del Golfo de M\'exico'' project
was funded by PEMEX contracts SAP-428217896, 428218855, and 428229851.
The quality control and post-processing of the CICESE--Pemex data
were carried out by Paula Garcia and Argelia Ronquillo.   Support
for this work was provided by the Gulf of Mexico Research Initiative
(PM, FJBV and MJO) as part of CARTHE and Secretar\'{\i}a de
Energ\'{\i}a (SENER)--Consejo Nacional de Ciencia y Tecnolog\'{\i}a
(CONACyT) grant 201441 (FJBV, MJO, JS and PPB) as part of the
Consorcio de Investigaci\'on del Golfo de M\'exico (CIGoM).

\section*{Author Contributions}

P.M., F.J.B.V., M.J.O.\ and G.F.\ performed the eigenvectors analysis
and the extraction of the attraction zones and the basins of
attraction. P.M. and F.J.B.V.\ prepared the figures. M.J.O.\ generated
the drifter trajectories database using public data which was
augmented with data obtained by P.G.B.\ and J.S.\ from Horizon
Marine Inc. All authors contributed to the interpretation of the
results and the writing of the manuscript.

\section*{Additional Information}

\noindent\textbf{Competing financial interests:} The authors declare
no competing financial interests.

\newpage

\end{document}